\documentclass[preprint,showpacs,preprintnumbers,amsmath,amssymb,superscriptaddress]{revtex4}

\usepackage[dvips]{epsfig}
\usepackage{dcolumn}
\usepackage{bm}

\begin{document}

\newcommand{\rpv}{RPV}                  
\newcommand{\mll}{$m_{ll'}$}   
\newcommand{\msnu}{$m_{\tilde{\nu}}$}   
\def\met{\mbox{${\hbox{$E$\kern-0.6em\lower-.1ex\hbox{/}}}_T$}} 
\newcommand{\ipb}{$\rm{pb}^{-1}$}       
\newcommand{\gevc}{GeV/$c$}             
\newcommand{\gevcc}{GeV/$c^2$}          

\title{Search for lepton flavor violating decays of a heavy neutral particle
       in $p\overline p$ collisions at $\sqrt{s}=1.8$ TeV}

\font\eightit=cmti8
\def\r#1{\ignorespaces $^{#1}$}
\hfilneg
\author{
\noindent
D.~Acosta,\r {14} T.~Affolder,\r 7 H.~Akimoto,\r {51}
M.G.~Albrow,\r {13} D.~Ambrose,\r {37}   
D.~Amidei,\r {28} K.~Anikeev,\r {27} J.~Antos,\r 1 
G.~Apollinari,\r {13} T.~Arisawa,\r {51}
% ...
}

\affiliation{
\r 1  {\eightit Institute of Physics, Academia Sinica, Taipei, Taiwan 11529, 
Republic of China} \\
\r 2  {\eightit Argonne National Laboratory, Argonne, Illinois 60439} \\
\r 3  {\eightit Istituto Nazionale di Fisica Nucleare, University of Bologna,
I-40127 Bologna, Italy} \\
\r 4  {\eightit Brandeis University, Waltham, Massachusetts 02254} \\
\r 5  {\eightit University of California at Davis, Davis, California  95616} \\
\r 6  {\eightit University of California at Los Angeles, Los 
Angeles, California  90024} \\ 
\r 7  {\eightit University of California at Santa Barbara, Santa Barbara, California 
93106} \\ 
\r 8 {\eightit Instituto de Fisica de Cantabria, CSIC-University of Cantabria, 
39005 Santander, Spain} \\
\r 9  {\eightit Carnegie Mellon University, Pittsburgh, Pennsylvania  15213} \\
\r {10} {\eightit Enrico Fermi Institute, University of Chicago, Chicago, 
Illinois 60637} \\
\r {11}  {\eightit Joint Institute for Nuclear Research, RU-141980 Dubna, Russia}
\\
\r {12} {\eightit Duke University, Durham, North Carolina  27708} \\
\r {13} {\eightit Fermi National Accelerator Laboratory, Batavia, Illinois 
60510} \\
\r {14} {\eightit University of Florida, Gainesville, Florida  32611} \\
\r {15} {\eightit Laboratori Nazionali di Frascati, Istituto Nazionale di Fisica
               Nucleare, I-00044 Frascati, Italy} \\
\r {16} {\eightit University of Geneva, CH-1211 Geneva 4, Switzerland} \\
\r {17} {\eightit Glasgow University, Glasgow G12 8QQ, United Kingdom}\\
\r {18} {\eightit Harvard University, Cambridge, Massachusetts 02138} \\
\r {19} {\eightit Hiroshima University, Higashi-Hiroshima 724, Japan} \\
\r {20} {\eightit University of Illinois, Urbana, Illinois 61801} \\
\r {21} {\eightit The Johns Hopkins University, Baltimore, Maryland 21218} \\
\r {22} {\eightit Institut f\"{u}r Experimentelle Kernphysik, 
Universit\"{a}t Karlsruhe, 76128 Karlsruhe, Germany} \\
\r {23} {\eightit High Energy Accelerator Research Organization (KEK), Tsukuba, 
Ibaraki 305, Japan} \\
\r {24} {\eightit Center for High Energy Physics: Kyungpook National
University, Taegu 702-701; Seoul National University, Seoul 151-742; and
SungKyunKwan University, Suwon 440-746; Korea} \\
\r {25} {\eightit Ernest Orlando Lawrence Berkeley National Laboratory, 
Berkeley, California 94720} \\
\r {26} {\eightit University College London, London WC1E 6BT, United Kingdom} \\
\r {27} {\eightit Massachusetts Institute of Technology, Cambridge,
Massachusetts  02139} \\   
\r {28} {\eightit University of Michigan, Ann Arbor, Michigan 48109} \\
\r {29} {\eightit Michigan State University, East Lansing, Michigan  48824} \\
\r {30} {\eightit Institution for Theoretical and Experimental Physics, ITEP,
Moscow 117259, Russia} \\
\r {31} {\eightit University of New Mexico, Albuquerque, New Mexico 87131} \\
\r {32} {\eightit Northwestern University, Evanston, Illinois  60208} \\
\r {33} {\eightit The Ohio State University, Columbus, Ohio  43210} \\
\r {34} {\eightit Osaka City University, Osaka 588, Japan} \\
\r {35} {\eightit University of Oxford, Oxford OX1 3RH, United Kingdom} \\
\r {36} {\eightit Universita di Padova, Istituto Nazionale di Fisica 
          Nucleare, Sezione di Padova, I-35131 Padova, Italy} \\
\r {37} {\eightit University of Pennsylvania, Philadelphia, 
        Pennsylvania 19104} \\   
\r {38} {\eightit Istituto Nazionale di Fisica Nucleare, University and Scuola
               Normale Superiore of Pisa, I-56100 Pisa, Italy} \\
\r {39} {\eightit University of Pittsburgh, Pittsburgh, Pennsylvania 15260} \\
\r {40} {\eightit Purdue University, West Lafayette, Indiana 47907} \\
\r {41} {\eightit University of Rochester, Rochester, New York 14627} \\
\r {42} {\eightit Rockefeller University, New York, New York 10021} \\
\r {43} {\eightit Instituto Nazionale de Fisica Nucleare, Sezione di Roma,
University di Roma I, ``La Sapienza," I-00185 Roma, Italy}\\
\r {44} {\eightit Rutgers University, Piscataway, New Jersey 08855} \\
\r {45} {\eightit Texas A\&M University, College Station, Texas 77843} \\
\r {46} {\eightit Texas Tech University, Lubbock, Texas 79409} \\
\r {47} {\eightit Institute of Particle Physics, University of Toronto, Toronto
M5S 1A7, Canada} \\
\r {48} {\eightit Istituto Nazionale di Fisica Nucleare, University of Trieste/\
Udine, Italy} \\
\r {49} {\eightit University of Tsukuba, Tsukuba, Ibaraki 305, Japan} \\
\r {50} {\eightit Tufts University, Medford, Massachusetts 02155} \\
\r {51} {\eightit Waseda University, Tokyo 169, Japan} \\
\r {52} {\eightit University of Wisconsin, Madison, Wisconsin 53706} \\
\r {53} {\eightit Yale University, New Haven, Connecticut 06520} \\
}

\collaboration{The CDF Collaboration}
\noaffiliation

\begin{abstract}
We report on a
search for a high mass, narrow width particle that 
decays directly to $e\mu$, $e\tau$, or $\mu\tau$.
We use approximately 110 \ipb\ of data collected with
the Collider Detector at Fermilab from 1992 to 1995.
No evidence of lepton flavor violating decays is found.
Limits are set on the production and decay of sneutrinos with R-parity 
violating interactions.
\end{abstract}


\maketitle

Particles that decay to $e\mu$, $e\tau$, or $\mu\tau$
occur in a number of extensions to the standard model.
Examples include 
Higgs bosons in models with multiple Higgs doublets 
\cite{lfvh99,lfvh00,lfvh02},
sneutrinos in supersymmetric models with R-parity violation (\rpv)
\cite{dehms,krsz,shwg}, 
horizontal gauge bosons \cite{bhss}, 
and $Z'$ bosons \cite{lfvz}.
In this paper we report results from a search based on final states
containing $e\mu$.
We are sensitive to $e\tau$ and $\mu\tau$ modes through
$\tau \to \mu \nu \nu$ and $\tau \to e\nu \nu$ 
respectively.
We analyze data from $p\overline p$ collisions
at center of mass energy $\sqrt{s} = 1.8$ TeV recorded with 
the Collider Detector at Fermilab (CDF) during the 1992 to 1995 Tevatron run.
The integrated luminosity is approximately 110 \ipb. 

The CDF detector has been described in detail elsewhere \cite{cdfdet}. 
This analysis makes use of several detector subsystems.
The position of $p\overline p$ collisions along the
beam line is measured in the vertex time projection chambers (VTX).
The central tracking chamber (CTC), 
located within the 1.4 T magnetic field of a 
superconducting solenoid, measures the momentum of charged particles. 
The transverse momentum resolution for muon and charged hadron tracks 
in the pseudorapidity interval $|\eta| < 1.1$ that are constrained to 
originate at the beamline is better than $0.1\% \times p_T$, 
where $p_T$ is measured in \gevc\ and is the momentum
component transverse to the beam line.
Sampling calorimeters surround the solenoid.
The central electromagnetic calorimeter (CEM) covers $|\eta| < 1.1$
and measures the energy of electromagnetic showers with a resolution of 
about $(13.7/\sqrt{E_T}\oplus 2)$\%,
where $\oplus$ means addition in quadrature
and the transverse energy $E_T$ is measured in GeV. 
The transverse energy is defined as $E \sin(\theta)$, with $E$ being 
the shower energy measured in the calorimeter and $\theta$ the polar angle 
of the energy flow, from the $p\overline p$ interaction vertex to the 
calorimeter deposition.
Within the CEM, proportional chambers (CES) measure the transverse shape of 
showers.
Drift chambers located outside the calorimeters detect muons in the 
region $|\eta| < 1$.
A measure of the energy carried by neutrinos escaping the detector
is the missing transverse energy \met,
calculated from the vector
sum of the energy depositions in the calorimeters and the
momentum  of muon tracks.

Events containing an electron and a muon are recorded by an assortment of 
single lepton, dilepton, and jet triggers \cite{cdfev}.
We select events that have an electron with $E_T > 20$ GeV, 
a muon with $p_T > 20$ \gevc, 
and a primary $p\overline p$ interaction vertex within 60 cm of the 
center of the detector.
The electron and muon must have opposite charges.
We identify electrons and muons using criteria that retain efficiency
for very high momentum particles \cite{zprime}.
Electrons must have 
a shower contained within the sensitive region of the CEM and have
a CTC track with $p_T > 13$ \gevc\ that matches the position of the shower
in the CES.
The CEM determines the electron energy. 
Requirements on the $p_T$ of the associated CTC track are kept loose 
because electrons can lose energy in the tracking volume due to bremsstrahlung. 
In cases in which an electron has 
$E_T < 100$ GeV or $p_T < 25$ \gevc\ 
there are additional requirements:
the lateral distribution of energy must be
consistent with an electromagnetic shower, and the ratio of energy measured
in the calorimeter to the momentum measured in the CTC must be less than four.
Electrons that appear to be one leg of a photon conversion --- because 
the second leg is found or because the CTC track is not confirmed by hits
in the VTX --- are removed.
Muons must have a track in the CTC that originates from the 
primary vertex and matches a track segment in the muon system. 
The energy along the muon path through the calorimeters must be consistent
with a minimum ionizing particle. 
Electrons and muons are both required to be isolated from other energy
deposition
in the calorimeter.
The combination of triggers is, within one standard deviation, 
fully efficient for events that satisfy the offline selection.

Backgrounds with two isolated high $p_T$ leptons arise from standard model
production of $Z/\gamma$, $WW$, $WZ$, $ZZ$, and $t\overline t$.
The cross sections for the first four of these processes are calculated at 
next-to-leading order using MCFM v1.0 \cite{mcfm} and MRST99 parton
distribution functions \cite{mrst99}.
The $t\overline t$ production cross section is taken from the parameterized 
formula of Ref. \cite{catani}
evaluated at 
the Tevatron average 
top quark mass of 174.3 GeV/$c^2$ \cite{pdg2000}.
For each of the five processes, 
the fraction of events expected to pass the offline selection
(the acceptance)
is calculated by Monte Carlo 
using Pythia $\rm{v}5.7$ \cite{pythia57}, 
the CDF detector simulation,
and the same analysis software used to select events in the data.
The efficiency of the lepton identification requirements in
the simulation is calibrated
using electrons and muons from $Z$ decays.
For each background process, the expected number of events is estimated to be
$\sigma_B A_B n_{ee} / (\sigma_{ee} A_{ee})$, 
where 
$\sigma_B$ is the cross section for the background process,
$A_B$ is the acceptance for the background process,
$n_{ee}$ is the number of $e^+e^-$ events 
observed in the data with a mass in the region of the $Z$ peak,
$\sigma_{ee}$ is the $Z/\gamma$ cross section times branching fraction
to $e^+e^-$, and $A_{ee}$ is the acceptance for $e^+e^-$.
By normalizing to $e^+e^-$ data
we eliminate uncertainties in the background level due to 
the integrated luminosity
and 
the $Z/\gamma$ cross section times branching fraction to leptons,
and we reduce the uncertainty from lepton efficiencies.
The $t\overline t$, $WW$, $WZ$, and $ZZ$ cross sections
are assigned uncertainies of 25\%, 11\%, 10\%, and 10\% respectively.
Other uncertainties are from
the detector model (6\%), 
Monte Carlo statistics (2--4\%), 
$Z\to ee$ statistics (2\%),
particle identification efficiencies (2\%),
and trigger efficiencies (1\%).
The remaining backgrounds
arise from particles produced in jets.
These include instrumental fakes and real leptons
from $b$- and $c$-quark decays.
All of them are denoted false leptons.
The false lepton backgrounds are estimated using
a method similar to that of Ref. \cite{d0wbr}, 
in which the probability for a false lepton to appear isolated
is measured in a control sample of dijet data.

\begin{figure}[hbt]
\psfig{file=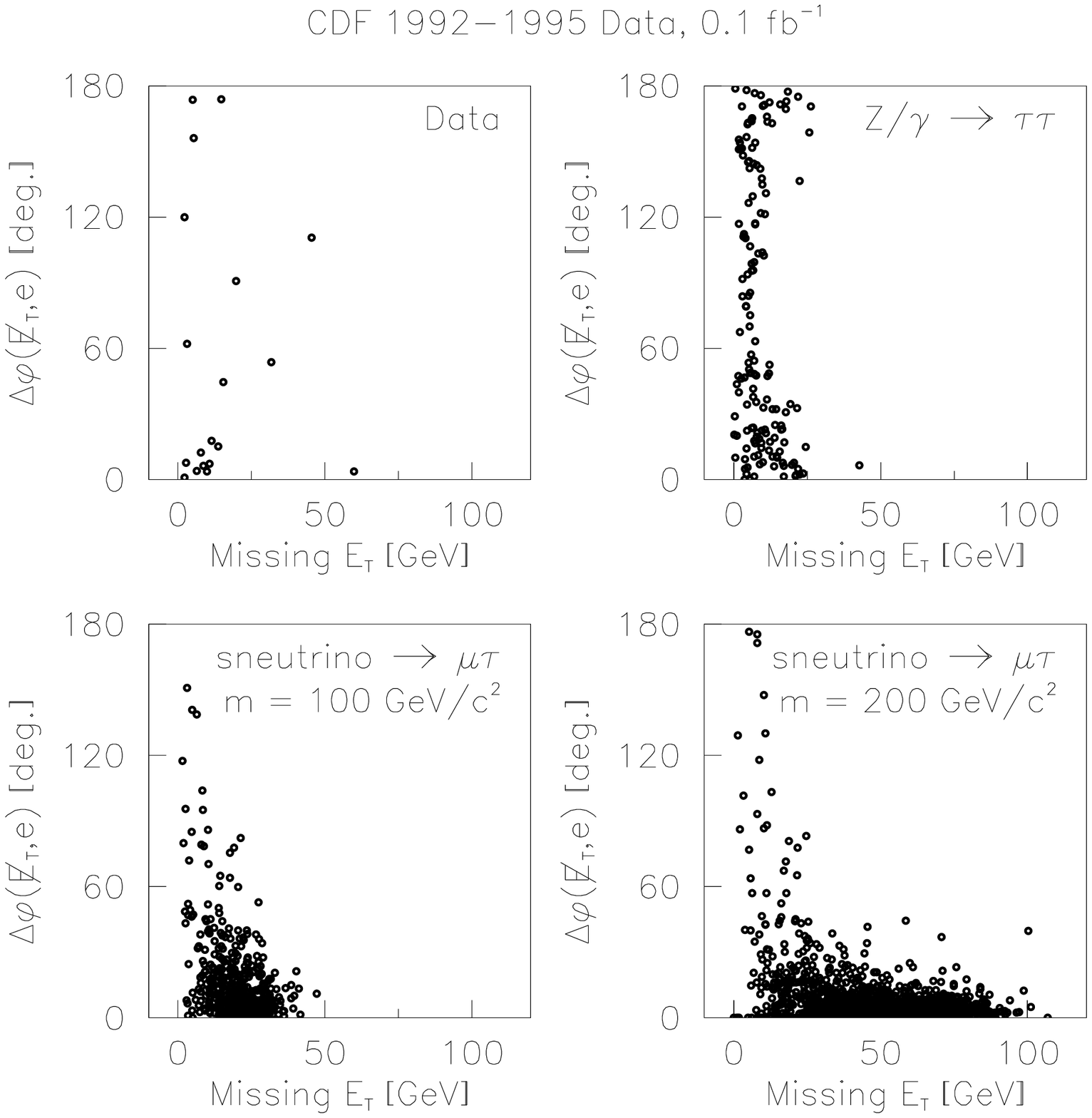,width=\linewidth}
\caption{\label{fig-phimete} 
  The angle in the plane transverse to the beamline between the direction 
  of the missing energy and the electron versus the missing transverse energy:
  data (upper left);
  simulated $Z/\gamma \to \tau\tau \to e\mu\nu\nu$, 
  the dominant background (upper right);
  and a hypothetical signal $\tilde{\nu} \to \mu\tau \to \mu e \nu$ for
  a sneutrino mass of 100 \gevcc\ (lower left) and 
  200 \gevcc\ (lower right).
  The requirement $\Delta\phi(e,\mu) > 120^\circ$ has already been imposed.
}
\end{figure}

In a $WW$ or $t \overline t$ event, 
the electron and the muon, originating from different $W$ bosons, 
have largely independent directions, 
spin correlations not withstanding.
To reduce these backgrounds,
we require that the angle between the electron and muon 
in the plane transverse to the beam
be at least 120 degrees.
This is almost fully efficient for two-body decays of a particle as heavy as
the $Z$ boson. 
For the $e\mu$ channel, there are no further requirements.
An $e\mu$ event is additionally classified as an $e\tau$ event 
if the angle between the \met\ and the muon,
$\Delta \phi (\mu,{\hbox{$E$\kern-0.6em\lower-.1ex\hbox{/}}}_T)$,
is less than 60 degrees. 
This eliminates events that are not consistent with $\tau \to \mu\nu\nu$,
since the tau is energetic enough that its decay products are nearly collinear.
Likewise, an $e\mu$ event is additionally classified as a $\mu\tau$ event if 
$\Delta \phi (e,{\hbox{$E$\kern-0.6em\lower-.1ex\hbox{/}}}_T) < 60$ degrees. 
The distribution of
$\Delta \phi (e,{\hbox{$E$\kern-0.6em\lower-.1ex\hbox{/}}}_T)$ versus \met\
is shown in Figure \ref{fig-phimete}. 
Since the electrons and muons are nearly back-to-back in $\phi$,
$\Delta \phi (\mu,{\hbox{$E$\kern-0.6em\lower-.1ex\hbox{/}}}_T)$ 
is approximately 
$\Delta \phi (e,{\hbox{$E$\kern-0.6em\lower-.1ex\hbox{/}}}_T) + 180$ degrees.

There are nineteen $e\mu$ candidates in the data.
Four of these are also $e\tau$ candidates, 
and twelve are also $\mu\tau$ candidates.
The contributions of the various backgrounds to each channel are listed 
in Table \ref{tab-bkg}.
The dominant source of background is $Z/\gamma \to \tau\tau$
where one tau decays to $e\nu\nu$ and the other to $\mu\nu\nu$.
The small contribution from $Z/\gamma \to \mu\mu$ is from cases in which
one muon is mistaken for an electron after radiating an energetic photon. 
\begin{table}[ht]
\begin{center}
\begin{tabular}{lrrrrrr}
\hline
\hline
\multicolumn{1}{l}{Source} &&
\multicolumn{1}{c}{$e\mu$} &&
\multicolumn{1}{c}{$e\tau$} &&
\multicolumn{1}{c}{$\mu\tau$} \\
\hline
$Z/\gamma \to \tau\tau$ && $13.91 \pm 0.99$ && $5.20 \pm 0.43$ && $6.48 \pm 0.59$ \\
\multicolumn{1}{l}{ $Z/\gamma \to \mu\mu$ } &&
\multicolumn{1}{r}{ $0.27 \pm 0.16$ } &&
\multicolumn{1}{c}{ 0 } &&
\multicolumn{1}{r}{ $0.27 \pm 0.16$ } \\
$WW$                    && $2.37 \pm 0.32$ && $0.42 \pm 0.07$ && $0.45 \pm 0.08$ \\
$WZ$                    && $0.13 \pm 0.02$ && $0.03 \pm 0.01$ && $0.04 \pm 0.01$ \\
$ZZ$                    && $0.024 \pm 0.004$ && $0.007 \pm 0.001$ && $0.007 \pm 0.002$ \\
$t\overline t$          && $1.34 \pm 0.36$ && $0.40 \pm 0.11$ && $0.35 \pm 0.10$ \\
False $e$               && $0.85 \pm 0.44$ && $-0.04 \pm 0.23$ && $0.96 \pm 0.32$ \\
False $\mu$             && $0.98 \pm 0.27$ && $-0.06 \pm 0.11$ && $1.07 \pm 0.25$ \\
\hline
Total Background        && $19.88 \pm 1.42$ && $5.95 \pm 0.55$ && $9.62 \pm 0.81$ \\
\hline
\multicolumn{1}{l}{Data} &&
\multicolumn{1}{c}{19} &&
\multicolumn{1}{c}{4} &&
\multicolumn{1}{c}{12} \\
\hline
\hline
\end{tabular}
\caption{The expected number of background events and the observed
         number of candidates in each channel.}
\label {tab-bkg}
\end{center}
\end{table}

\begin{figure}[hbt]
\psfig{file=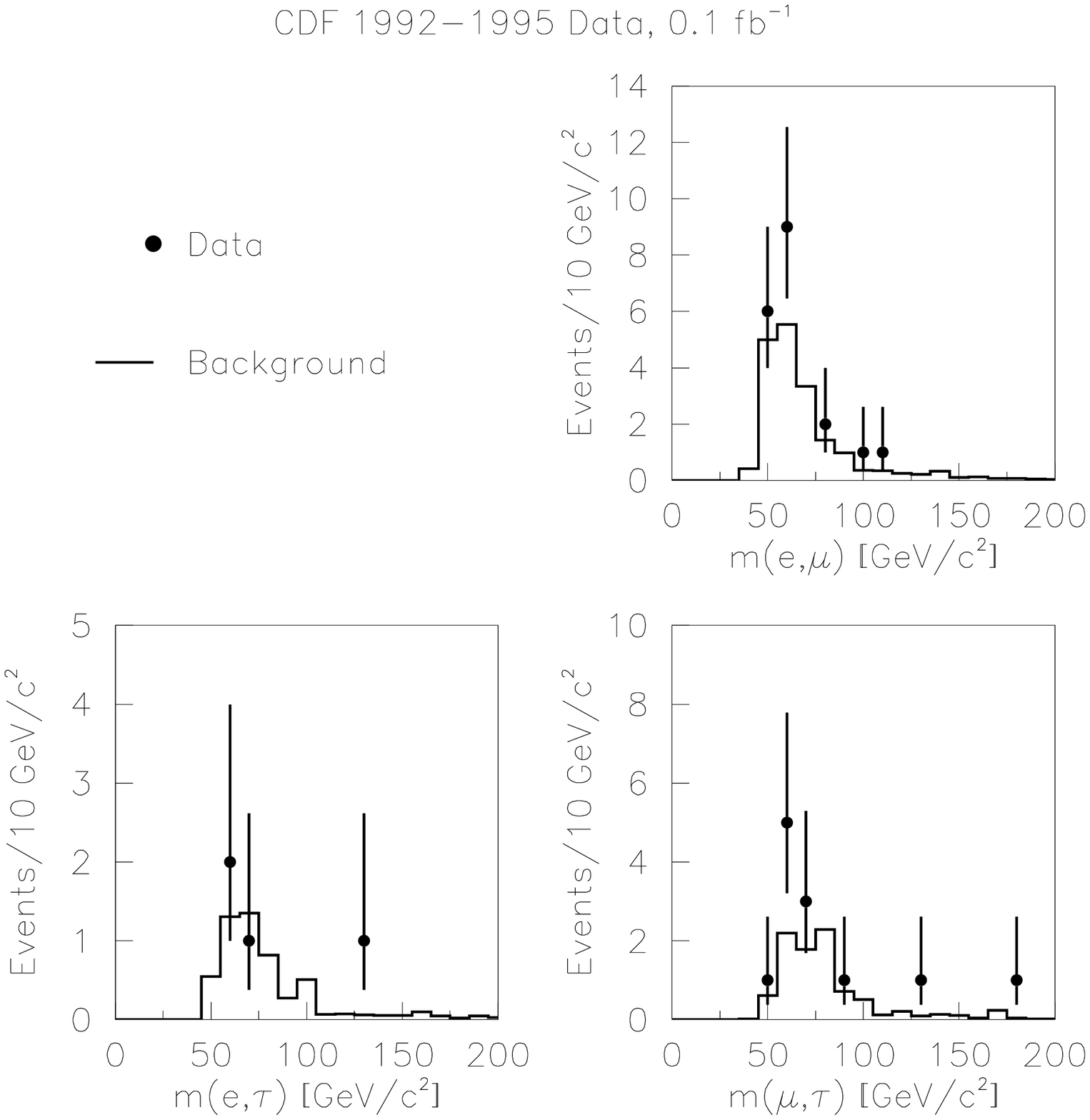,width=\linewidth}
\caption{\label{fig-msnu} 
 The reconstructed mass of the lepton pairs.
 The points show
 the nineteen $e\mu$ candidates (upper),
 the four $e\tau$ candidates (lower left),
 and the twelve $\mu\tau$ candidates (lower right).
 The histograms show the total background in each sample.
 The $e\tau$ and $\mu\tau$ samples have no events in common;
 both are subsets of the $e\mu$ sample.
 \met\ is used in computing $\tau$ momentum, as explained in the text.
}
\end{figure}

The distributions of the lepton pair masses \mll\
are shown in Figure \ref{fig-msnu}.
For the $e\tau$ and $\mu\tau$ hypotheses, \mll\ is
calculated assuming that the tau momentum components are
$  p_x^{\tau}  =  p_x^l + {E\!\!\!\!/}_x $,
$  p_y^{\tau}  =  p_y^l + {E\!\!\!\!/}_y $, and
$  p_z^{\tau}  =  p_z^l \times (1 + {E\!\!\!\!/}_T / p_T^l) $,
where $p^l$ is the momentum of the electron or muon to which the tau
is assumed to have decayed \cite{lfvh00,lfvh02}. 
The data show no indication of a signal peak.
The distribution of observed events is compared to the expected background
shape using a Kolmogorov test.
The test statistic probability is 
19\%, 65\%, and 74\% for $e\mu$, $e\tau$, and $\mu\tau$ respectively,
consistent with the absence of a signal.

As a specific signal model,
we consider the process 
$d \overline d \to \tilde{\nu} \to ll'$ 
mediated by \rpv\ interactions
\cite{dehms,krsz,shwg}.
The \rpv\ sneutrino couplings allowed in the supersymmetric Lagrangian are
$ [\lambda_{ijk} ( \tilde{\nu}_L^j \overline{e}_R^k e_L^i
                 -\tilde{\nu}_L^i \overline{e}_R^j e_L^k)
 + \lambda'_{ijk} \tilde{\nu}_L^i \overline{d}_R^k d_L^j]
 + \rm{h.c.}$\,
where the indices $i$, $j$, and $k$ label generations; $\lambda_{ijk}$ is
non-zero only for $i<j$; and mixing is ignored \cite{bgh,krsz}. 
Constraints on the coupling strengths have been derived from
measurements of low energy processes \cite{bgh, add}.
The upper bound on the coupling that mediates 
$d \overline d \to \tilde{\nu}_{\tau}$ is
$\lambda'_{311} < 0.11 \times \frac{m_{\tilde{d}_R}} {100 \,\rm{GeV}}$.
The limit is slightly stronger for $\lambda'_{211}$ and much
stronger for $\lambda'_{111}$.  
Of the couplings that contribute to $\tilde{\nu}_{\mu,\tau} \to ll'$ 
those with the loosest bounds are
$\lambda_{132} < 0.062 \times \frac{m_{\tilde{\mu}_R}} {100 \,\rm{GeV}}$,
$\lambda_{231} < 0.070 \times \frac{m_{\tilde{e}_R}} {100 \,\rm{GeV}}$,
$\lambda_{233} < 0.070 \times \frac{m_{\tilde{\tau}_R}} {100 \,\rm{GeV}}$,
and
$\lambda_{122} < 0.049 \times \frac{m_{\tilde{\mu}_R}} {100 \,\rm{GeV}}$.
Limits on $\mu$--$e$ conversion in nuclei severely restrict the 
$\lambda\lambda'$ products that contribute to the $e\mu$ channel,
for example $|\lambda_{231} \lambda_{311}'| < 4.1 \times 10^{-9}$ assuming
sparticle masses of 100 \gevcc\
\cite{faessler}.
Searches at the CERN $e^+e^-$ collider rule out
sneutrino masses $m_{\tilde{\nu}} < 86$ \gevcc\
if any one (but only one) of the $\lambda$ constants is non-zero \cite{aleph}.
Previous CDF searches have examined scenarios with 
$\lambda'_{121}$ \cite{cdf121}
and $\lambda'_{333}$ \cite{cdf333} non-zero.

We simulate the signal process
by generating events using the heavy higgs ($H'$) process in Pythia.
The $H'$ decay table is modified to include each of the 
$e\mu$, $e\tau$, and $\mu\tau$ modes in turn
while all other decay channels are switched off.
All initial states except for $d\overline{d}$ are inhibited.
Events are generated for nine particle masses \msnu\ between
50 and 800 \gevcc.
The events are passed through the CDF detector simulation
and the analysis software used for the data.
For each \msnu\ and each decay mode, the mean and rms \mll\ 
is computed.
The rms widths of the $e\mu$, $e\tau$, and $\mu\tau$ mass distributions
are 3, 6, and 7 \gevcc\ respectively
for $m_{\tilde{\nu}} = 100$ \gevcc.
For $m_{\tilde{\nu}} = 400$ \gevcc\ they are
30, 14, and 60 \gevcc.
The broadening with \msnu\ is
due to the 
muon $p_T$ resolution and, to a lesser degree, the electron energy resolution.
At low \msnu\ the resolution for the $e\tau$ and $\mu\tau$ modes is poorer than 
for $e\mu$ due to the inclusion of \met.
The $e\tau$ resolution degrades more slowly than the others because the
muons from tau decays have lower momenta.

To study the full range of \msnu\ without gaps in
which a signal might hide,
we count events within three standard deviations of the mean \mll\
for a sequence of \msnu\ values starting at 50 \gevcc\ with step size
from the $i$th to the ($i+1$)th value equal to one tenth of the rms of the
\mll\ distribution at the $i$th point.
The mean and rms of the \mll\ distribution for each \msnu\ is
linearly interpolated between values at the generated \msnu\ points.
A three standard deviation window provides good statistical sensitivity
while incurring little dependence on any possible systematic errors in the
width or mean.

\begin{figure}[hbt]
\psfig{file=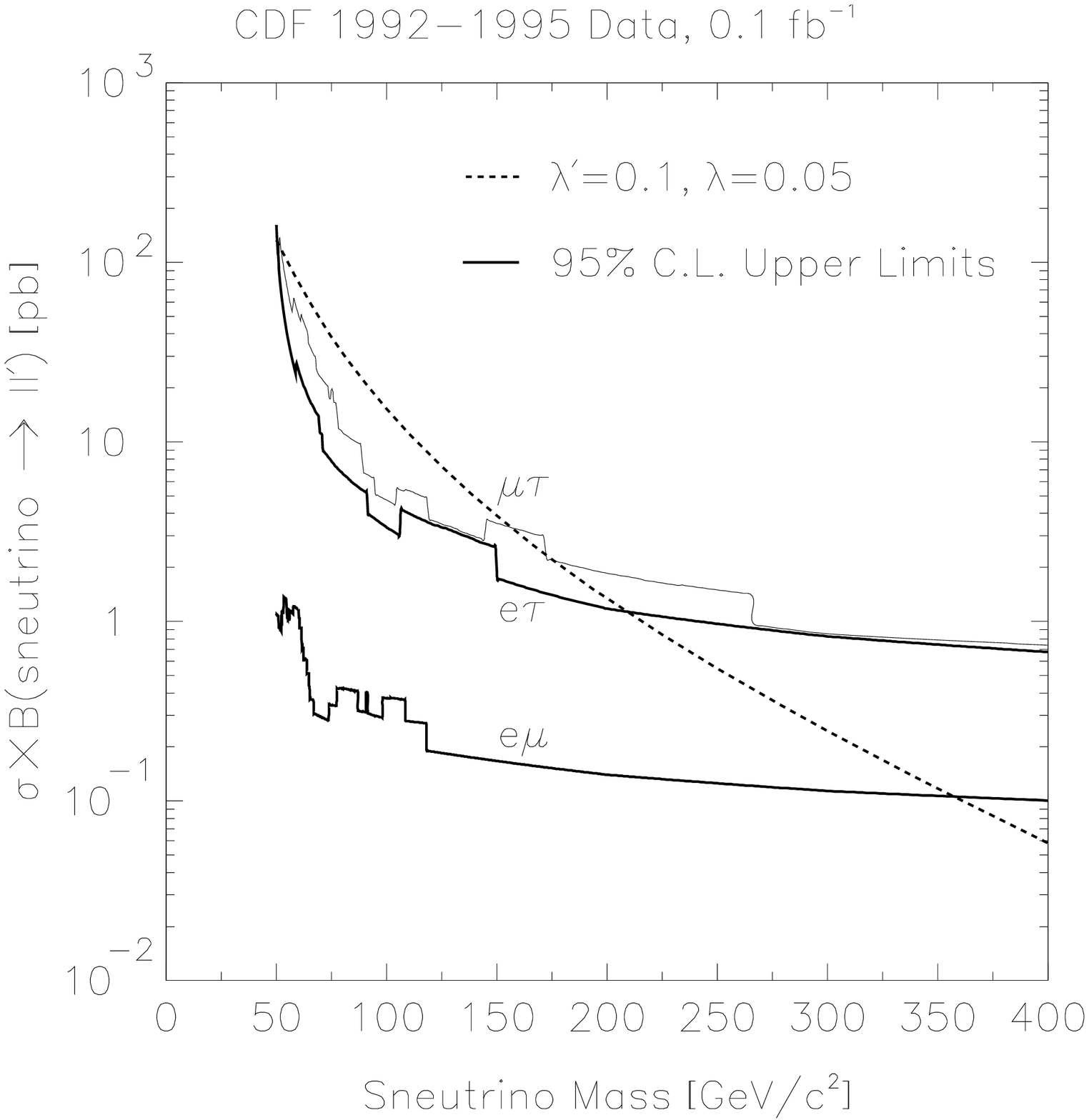,width=\linewidth}
\caption{\label{fig-limit} 
95\% C.L. upper limits on 
$\sigma \times B(\tilde{\nu} \to e\mu)$,
$\sigma \times B(\tilde{\nu} \to e\tau)$,
and $\sigma \times B(\tilde{\nu} \to \mu\tau)$
as a function of sneutrino mass, together with the next-to-leading order
cross section for the reference parameters.
}
\end{figure}

For each decay mode 
we derive 
a 95\% C.L. upper limit \cite{pdg1996}
on $\sigma \times B$ using the
number of events, expected background, and acceptance 
in the mass window corresponding to each \msnu\ point. 
Figure \ref{fig-limit} shows the limits 
as a function of \msnu.
The limits on 
$\sigma \times B(\tilde{\nu} \to e\tau)$ and
$\sigma \times B(\tilde{\nu} \to \mu\tau)$
are higher than the limit on
$\sigma \times B(\tilde{\nu} \to e\mu)$
because of the tau branching ratio to leptons and, particularly
at low \msnu, because the leptons from tau decays tend to fall below
the 20 \gevc\ $p_T$ threshold.

As a benchmark, Figure \ref{fig-limit} also shows 
the theoretical cross section times branching fraction for
$d\overline{d}\to\tilde{\nu}_{\tau}\to e^+\mu^-$
plus
$d\overline{d}\to\overline{\tilde{\nu}}_{\tau}\to e^-\mu^+$
as a function of \msnu\ in the case
$\lambda_{311}' = 0.1$ and $\lambda_{132} = 0.05$.
The curve is obtained using
next-to-leading order values of
$\sigma(d\overline d \to \tilde{\nu})$ 
for $\lambda' = 0.01$ \cite{nlo18,nlo2}, 
which we scale by $(\lambda'/0.01)^2$.
$B(\tilde{\nu} \to e\mu)$ is calculated assuming that
weak decays of the sneutrino are kinematically forbidden
and that the only non-zero \rpv\ couplings 
are $\lambda_{311}'$ and $\lambda_{132}$.

In conclusion, we find no evidence for new particles with lepton
flavor violating decays. We set limits on the cross section 
for single sneutrino production times 
the branching fractions
$B(\tilde{\nu} \to e\mu$),
$B(\tilde{\nu} \to e\tau$), and
$B(\tilde{\nu} \to \mu\tau$)
as a function of the sneutrino mass.
For a sneutrino mass of 200 \gevcc, 
the 95\% C.L. upper limits on $\sigma \times B$ are
0.14 pb, 1.2 pb, and 1.9 pb for the $e\mu$, $e\tau$, and $\mu\tau$
modes respectively.

We thank the Fermilab staff and the technical staffs of the 
participating institutions for their vital contributions.  
We are grateful to Debajyoti Choudhury and Swapan Majhi for providing
NLO sneutrino cross sections for $\sqrt{s} = 1.8$ TeV.
This work was supported by 
the U.S. Department of Energy and National Science Foundation; 
the Italian Istituto Nazionale di Fisica Nucleare; 
the Ministry of Education, Culture, Sports, Science, and Technology of Japan; 
the Natural Sciences and Engineering Research Council of Canada; 
the National Science Council of the Republic of China; 
the Swiss National Science Foundation; 
the A.P. Sloan Foundation; 
the Bundesministerium fuer Bildung und Forschung, Germany; 
and the Korea Science and Engineering Foundation (KoSEF); 
the Korea Research Foundation; 
and the Comision Interministerial de Ciencia y Tecnologia, Spain.

\end{document}